\documentstyle[12pt,epsfig,graphics,cite,amssymb,amsmath,bm]{article}
\begin{document}\bibliographystyle{plain}\begin{titlepage}
\renewcommand{\thefootnote}{\fnsymbol{footnote}}\hfill
\begin{tabular}{l}HEPHY-PUB 964/16\\
March 2016\end{tabular}\\[2cm]\Large\begin{center}{\bf
INSTANTANEOUS BETHE--SALPETER KERNEL FOR THE LIGHTEST PSEUDOSCALAR
MESONS}\\[1cm]\large{\bf Wolfgang LUCHA\footnote[1]{\normalsize\
{\em E-mail address\/}:
wolfgang.lucha@oeaw.ac.at}}\\[.3cm]\normalsize Institute for High
Energy Physics,\\Austrian Academy of Sciences,\\Nikolsdorfergasse
18, A-1050 Vienna, Austria\\[1cm]\large{\bf Franz
F.~SCH\"OBERL\footnote[2]{\normalsize\ {\em E-mail address\/}:
franz.schoeberl@univie.ac.at}}\\[.3cm]\normalsize Faculty of
Physics, University of Vienna,\\Boltzmanngasse 5, A-1090 Vienna,
Austria\\[2cm]{\normalsize\bf Abstract}\end{center}\normalsize

\noindent Starting from a phenomenologically successful, numerical
solution of the Dyson--Schwinger equation that governs the quark
propagator, we reconstruct in detail the \emph{interaction
kernel\/} that has to enter the instantaneous approximation to the
Bethe--Salpeter equation to allow us to describe the
\emph{lightest\/} pseudoscalar mesons as quark--antiquark bound
states exhibiting the (almost) masslessness necessary for them to
be interpretable as the (pseudo) Goldstone bosons related to the
spontaneous chiral symmetry breaking of quantum
chromodynamics.\vspace{3ex}

\renewcommand{\thefootnote}{\arabic{footnote}}\end{titlepage}

\section{Introduction}As part of the generally accepted basis (for
purely historical reasons frequently still dubbed as standard
``model'') of theoretical elementary particle physics,
\emph{quantum chromodynamics\/} (QCD), the relativistic quantum
field theory that describes all \emph{strong interactions\/}
between its colour-carrying fundamental degrees of freedom ---
that is, the quarks and the gluons --- is (strongly believed to
be) characterized by two essential properties: colour confinement
--- expressing the empirically established fact of non-observation
of isolated coloured particles in nature --- and dynamical
breakdown of chiral symmetry --- manifested by the emergence of
nonvanishing masses of nonperturbative origin for, at Lagrangian
level, massless quarks.

In principle, relativistic quantum field theory describes bound
states of its fundamental degrees of freedom by Bethe--Salpeter
amplitudes obtained by solution of the homogeneous Bethe--Salpeter
equation \cite{BSE}. For technical or practical reasons, the
latter is frequently used in some instantaneous limit
\cite{WL05:IBSEWEP}, which --- upon assuming, for each bound-state
constituent, free propagation with constant effective mass ---
finally collapses to Salpeter's equation \cite{SE}.

QCD enables to regard light pseudoscalar mesons as bound states of
quarks and gluons, or as pseudo-Goldstone bosons whose presence is
demanded by the Goldstone theorem as a consequence of
\emph{spontaneous\/} chiral symmetry breaking. Recently, we
embarked on the quest for reconciling both views by a
well-tailored instantaneous Bethe--Salpeter formalism
\cite{WL13,WL15,WL16:ARP}.

Instead of attempting to solve a given bound-state equation in use
straightforwardly, its exact solutions may be likewise determined,
along a more indirect route and only with a bit of luck, by
application of sufficiently sophisticated inversion techniques
which establish the \emph{rigorous\/} relation between the
relevant interactions and the wave functions (in our case, the
Bethe--Salpeter amplitudes) that characterize the features of the
bound~states~under study. In Ref.~\cite{WL13}, we demonstrated the
feasibility of this inversion concept for the easier-to-handle
example of the projection of the Salpeter equation to positive
energies of the~bound~quarks.

All symmetries of a quantum field theory are reflected by its
Ward--Takahashi identities rephrasing the implications of such
symmetries at the level of the theory's Green functions. In
Ref.~\cite{WL15}, we employed an ensued relation between
flavour-nonsinglet pseudoscalar-meson Bethe--Salpeter amplitude
and quark propagator to implement the proper ultraviolet limit.

General insights, rooting in axiomatic foundations of quantum
field theory and encoded in the axiom of reflection positivity,
impose constraints on the propagators of those degrees of freedom
of any quantum field theory that are bound to be subject to colour
confinement. In Ref.~\cite{WL16:ARP}, we added constraints on the
momentum dependence of the quark mass function, implied by the
nonexistence of a K\"all\'en--Lehmann representation for the quark
propagator.

Within our instantaneous Bethe--Salpeter approach, the effective
interaction governing bound states is fixed by the quark
couplings' Lorentz nature and a central potential $V(|\bm{x}|).$
In this study, we pin down $V(|\bm{x}|)$ from a numerical solution
of the quark Dyson--Schwinger equation exploiting the at present
presumably most popular model for the necessary input.

The outline of this paper is as follows. In Sec.~\ref{Sec:MIBSE},
we sketch, only to the extent absolutely necessary for any
self-contained presentation, the established description of bound
states of fermion and antifermion by means of the Bethe--Salpeter
framework in instantaneous~limit. In Sec.~\ref{Sec:DSBEC}, we
briefly recall the interplay between Bethe--Salpeter amplitude and
quark mass function, established by the axial-vector
Ward--Takahashi identity for QCD by relating the solution of the
\emph{Bethe--Salpeter equation\/} and that of the quark's
\emph{Dyson--Schwinger~equation\/}. In Sec.~\ref{Sec:RLT}, we
recall our favourite model for the latter. In Sec.~\ref{Sec:Hum},
we infer $V(|\bm{x}|)$ as a~function of the (at this stage) only
free parameter, the effective quark mass. In Sec.~\ref{Sec:SCIO},
we discuss our observations. (For convenience of notation, we
adopt natural units throughout: $\hbar=c=1.$)

\section{Mesons by Instantaneous Bethe--Salpeter Equation}
\label{Sec:MIBSE}In quantum field theory, the Bethe--Salpeter
approach describes a two-particle bound state in terms of its
Bethe--Salpeter amplitude $\Phi(p,P),$ depending on both relative
momentum $p$ and total momentum $P$ of its bound-state
constituents (suppressing indices for simplicity). The homogeneous
Bethe--Salpeter equation \cite{BSE} controlling $\Phi(p,P)$
reduces, for bound-state constituents experiencing exclusively
instantaneous interactions and having propagators of a
sufficiently trivial dependence on $p_0,$ to the instantaneous
Bethe--Salpeter equation \cite{WL05:IBSEWEP}~for\begin{equation}
\phi(\bm{p})\equiv\frac{1}{2\pi}\int{\rm d}p_0\,\Phi(p)\
,\label{Eq:SA}\end{equation}the (equal-time) Salpeter amplitude.
If the latter requirement is satisfied by approximating the
propagators by their free forms involving effective masses, the
bound-state equation for $\phi(\bm{p})$ simplifies further to the
Salpeter equation \cite{SE}. For bound states composed of~a
fermion and an antifermion (with masses $m_{1,2}$ and momenta
$\bm{p}_{1,2},$ respectively), that equation
reads\begin{align}\phi(\bm{p})&=\int\frac{{\rm d}^3q}{(2\pi)^3}
\left(\frac{\Lambda_1^+(\bm{p}_1)\,\gamma_0\,
[K(\bm{p},\bm{q})\,\phi(\bm{q})]\,\gamma_0\,\Lambda_2^-(\bm{p}_2)}
{P_0-E_1(\bm{p}_1)-E_2(\bm{p}_2)}\right.\nonumber\\[1ex]
&\hspace{11.11ex}\left.-\frac{\Lambda_1^-(\bm{p}_1)\,\gamma_0\,
[K(\bm{p},\bm{q})\,\phi(\bm{q})]\,\gamma_0\,\Lambda_2^+(\bm{p}_2)}
{P_0+E_1(\bm{p}_1)+E_2(\bm{p}_2)}\right),\label{Eq:SE}\end{align}
with free energy and positive/negative energy projectors of the
two constituents defined~by$$E_i(\bm{p})\equiv
\sqrt{\bm{p}^2+m_i^2}\ ,\qquad\Lambda_i^\pm(\bm{p})
\equiv\frac{E_i(\bm{p})\pm \gamma_0\,(\bm{\gamma}\cdot\bm{p}+m_i)}
{2\,E_i(\bm{p})}\ ,\qquad i=1,2\ .$$Its instantaneous
Bethe--Salpeter kernel $K(\bm{p},\bm{q})$ subsumes Lorentz nature
and momentum dependence of the effective interactions taking place
between the bound-state constituents: the former by a set of
(generalized) Dirac matrices $\Gamma_i$ ($i=1,2),$ the latter by
corresponding Lorentz-scalar potential functions
$V_\Gamma(\bm{p},\bm{q}).$ For identical Lorentz structures of the
effective couplings of fermion and antifermion, i.e.,
$\Gamma_1=\Gamma_2=\Gamma,$ the action of $K(\bm{p},\bm{q})$ on
$\phi(\bm{p})$~reads$$[K(\bm{p},\bm{q})\,\phi(\bm{q})]=
\sum_\Gamma V_\Gamma(\bm{p},\bm{q})\,\Gamma\,\phi(\bm{q})\,\Gamma\
.$$Following Refs.~\cite{WL07a,WL07b,WL07c}, we rely on the
postulate of \emph{Fierz symmetry\/} of $K(\bm{p},\bm{q})$ by
choosing as its sole Lorentz structure (whence $V(\bm{p},\bm{q})$
no longer needs an index) the linear combination
$$\Gamma\otimes\Gamma=\frac{1}{2}\,(\gamma_\mu\otimes\gamma^\mu+
\gamma_5\otimes\gamma_5-1\otimes1)\ .$$

The $\Lambda^+\otimes\Lambda^-+\Lambda^-\otimes\Lambda^+$
projector structure of the right-hand side of Eq.~(\ref{Eq:SE})
implies~that, in a Dirac-space basis of 16 complex $4\times4$
matrices, the most general solution $\phi(\bm{p})$ has~eight
independent scalar component functions. Out of these, ignoring
flavour violation by letting $m_1=m_2=m,$ just two, henceforth
called $\varphi_i(\bm{p}),$ $i=1,2,$ correspond to spin-singlet
bound states of two spin-$\frac{1}{2}$ (anti-)fermions, such as
pseudoscalar mesons \cite{LOVW}. For these,~$\phi(\bm{p})$~reads
\begin{equation}\phi(\bm{p})=\left[\varphi_1(\bm{p})\,
\frac{\gamma_0\,(\bm{\gamma}\cdot\bm{p}+m)}{E(\bm{p})}
+\varphi_2(\bm{p})\right]\gamma_5\ .\label{Eq:PSA}\end{equation}

For \emph{spherically symmetric convolution-type\/} kernels
$K(\bm{p},\bm{q}),$ i.e., $V(\bm{p},\bm{q})=V((\bm{p}-\bm{q})^2),$
the Salpeter equation (\ref{Eq:SE}) simplifies to a set of
\emph{coupled\/} equations for the radial factors~$\varphi_i(p),$
$i=1,2,\dots,$ of the independent Salpeter components. Therein,
all interactions are encoded in form of configuration-space
central potentials, here generically denoted by $V(r),$
$r\equiv|\bm{x}|.$\pagebreak

Under all these assumptions, the Salpeter equation (\ref{Eq:SE})
can be shown to be equivalent to a system of two radial equations
\cite{WL07b}, an integral equation and a relation of algebraic
nature, determining mass eigenvalue $\widehat M$ and radial
Salpeter components $\varphi_{1,2}(p)$ of the~bound state:
\begin{align}&2\,E(p)\,\varphi_2(p)+2\int_0^\infty\frac{{\rm
d}q\,q^2}{(2\pi)^2}\,V(p,q)\,\varphi_2(q)=\widehat
M\,\varphi_1(p)\ ,\nonumber\\&2\,E(p)\,\varphi_1(p)=\widehat
M\,\varphi_2(p)\ ,\qquad E(p)\equiv\sqrt{p^2+m^2}\ ,\qquad
p\equiv|\bm{p}|\ .\label{Eq:5ess}\end{align}The potential $V(r)$
enters the first of these relations in form of its Fourier--Bessel
transform$$V(p,q)\equiv\frac{8\pi}{p\,q}\int_0^\infty{\rm d}r
\sin(p\,r)\sin(q\,r)\,V(r)\ ,\qquad q\equiv|\bm{q}|\ .$$

Within the context of the present kind of problem, inversion
simply means to determine the underlying interaction entering into
the generic type of equation of motion under~study by
\emph{postulating\/} one's favoured set of solutions. We take into
account the Goldstone nature~of the pseudoscalar bound states to
be described by Eq.~(\ref{Eq:5ess}) by requiring their mass
eigenvalue $\widehat M$ to vanish: $\widehat M=0.$ At this point
in our space of solutions, the two dynamical equations~in the set
(\ref{Eq:5ess}) decouple. The second relation implies
$\varphi_1(p)\equiv0,$ whence the Salpeter amplitude
(\ref{Eq:PSA}) simplifies to $\phi(\bm{p})=\varphi_2(\bm{p})\,
\gamma_5,$ and the first equation provides the sole component
$\varphi_2(\bm{p})$:\begin{equation}E(p)\,\varphi_2(p)+
\int_0^\infty\frac{{\rm d}q\,q^2}{(2\pi)^2}\,V(p,q)\,\varphi_2(q)
=0\ .\label{Eq:REE}\end{equation}With the Fourier--Bessel
transforms of the amplitude $\varphi_2(p)$ and the kinetic term
$E(p)\,\varphi_2(p)$\begin{align*}&\varphi(r)\equiv
\sqrt\frac{2}{\pi}\,\frac{1}{r} \int_0^\infty{\rm d}p\,p\sin(p\,r)
\,\varphi_2(p)\ ,\qquad T(r)\equiv\sqrt\frac{2}{\pi}\,\frac{1}{r}
\int_0^\infty{\rm d}p\,p\sin(p\,r)\,E(p)\,\varphi_2(p)\
,\end{align*}transforming Eq.~(\ref{Eq:REE}) to configuration
space enables the straightforward extraction of $V(r)$:
\begin{equation}T(r)+V(r)\,\varphi(r)=0\qquad\Longleftrightarrow
\qquad V(r)=-T(r)/\varphi(r)\ .\label{Eq:Po}\end{equation}In this
representation, it becomes plain that Eq.~(\ref{Eq:REE}) is just a
spinless Salpeter equation~\cite{WL-SSE}.

\section{The Dyson--Schwinger--Bethe--Salpeter Conspiracy}
\label{Sec:DSBEC}As an immediate consequence of the (global or
local) symmetries of a quantum field theory, Ward--Takahashi
identities relate differing $n$-point functions, e.g.,
\emph{propagators and vertices}. In the chiral limit, the
renormalized axial-vector QCD Ward--Takahashi identity relates the
dressed \emph{quark propagator\/} $S(p)$ to the
\emph{Bethe--Salpeter solution\/} $\Phi(p,0)$ for
flavour-nonsinglet pseudoscalar mesons: among others, a propagator
function basically determining the quark mass function in the
former to the \emph{dominant\/} ``Dirac'' component of the latter
\cite{PM97a,PM97b}. More precisely, if the two Lorentz-scalar
functions characterizing the propagator of the~quark are chosen to
be its mass function, $M(p^2),$ and its wave-function
renormalization factor,~$Z(p^2),$ $$S(p)=\frac{{\rm i}\,Z(p^2)}
{\not\!p-M(p^2)+{\rm i}\,\varepsilon}\ ,\qquad\not\!p\equiv
p^\mu\,\gamma_\mu\ ,\qquad\varepsilon\downarrow0\ ,$$and if the
--- comparatively low --- influence of the latter function is
neglected, \emph{in the~chiral limit\/} this Bethe--Salpeter
amplitude $\Phi(\underline{k},0)$ is found \cite{WL15} to be
related to $M(\underline{k}^2)$ according~to\begin{equation}
\displaystyle\Phi(\underline{k},0)\propto\frac{M(\underline{k}^2)}
{\underline{k}^2+M^2(\underline{k}^2)}\,\gamma_5+\mbox{subleading
contributions}\ ;\label{Eq:SP}\end{equation}here and below,
vectors in Euclidean space, where Eq.~(\ref{Eq:SP}) has been
derived, are underlined.

\section{Quark Propagator via Dyson--Schwinger Equations}
\label{Sec:RLT}By virtue of the relationship established by
Eq.~(\ref{Eq:SP}), the (approximate) pointwise behaviour of the
momentum-space Bethe--Salpeter amplitude for the massless
pseudoscalar mesons in the center-of-momentum frame can be
extracted from the quark mass function entering the quark
propagator. This two-point Green function, in turn, can be derived
as solution of the Dyson--Schwinger equation for the quark
propagator (sometimes dubbed as gap~equation).

With sufficient information about the Bethe--Salpeter amplitude
for the light mesons at our disposal, it is a (more or less)
straightforward enterprise to chart the sought interaction
potential: The Salpeter amplitude of any mesonic bound states
under consideration follows by application of its definition
(\ref{Eq:SA}). Then, the underlying potential may be read off from
the configuration-space representation of the effective equation
of motion according to~Eq.~(\ref{Eq:Po}).

Now, the Dyson--Schwinger equations constitute a countable
infinity of coupled integral equations for the infinite set of
$n$-point Green functions of a quantum field theory. However,
every member of this hierarchy of relations connects $n$-point
Green functions of different $n$: each Dyson--Schwinger equation
requires as input the solution of, at least, one higher-order
Dyson--Schwinger equation. Thus, the definition of a
\emph{tractable} problem is only possible by a truncation of this
infinite tower, to a finite set of equations for the Green
functions of low $n,$ with any required higher-$n$ input Green
function modelled by phenomenological~reasoning.

For obvious reasons, one's truncation of choice should best
respect all the symmetries of the quantum field theory expressed
by its Ward--Takahashi identities: The ``rainbow-ladder
truncation'' is defined by a tree-level quark--gluon vertex, a
Bethe--Salpeter kernel in ladder approximation relying on
single-gluon exchange, and a free gluon propagator. Nonetheless,
the desired conformity of the resulting predictions with
constraints imposed by experiment is, on the other hand, ensured
by effective coupling functions which substitute the square of the
strong couplings and have been composed such as to incorporate any
necessary~feature. This truncation preserves, at least, the
crucial QCD axial-vector Ward--Takahashi identity.

For the purposes of our present investigation, we will take
advantage of the findings of a study \cite{PM97b} adopting a
renormalization-group-improved truncation model devised such that
the effective coupling strength that defines a particular
rainbow-ladder truncation exhibits two decisive features: a
distinctive enhancement in the infrared region
($\underline{k}^2\to0$), motivated by the apparent behaviour of
the gluon propagator, and a decay similar to the perturbative
behaviour of the strong fine-structure coupling of QCD in the
ultraviolet~region ($\underline{k}^2\to\infty$).

Recognizing the importance of taking into account --- in the
course of application~of~our relation (\ref{Eq:SP}) --- the
behaviour of the quark mass function in the limit
$\underline{k}^2\to\infty$ as correctly as possible \cite{WL15},
we should capture a $\underline{k}^2$ interval which extends to
$\underline{k}^2$ values as large as~available. Hence, we utilize
the quark mass function $M(\underline{k}^2)$ in the form presented
in Fig.~2 of Ref.~\cite{PM97b}.

\section{Interaction Potentials from Quark Mass Functions}
\label{Sec:Hum}After these preliminaries, the potential $V(r)$
enabling the Salpeter equation (\ref{Eq:SE}) to describe massless
pseudoscalar solutions can be derived by an unspectacular sequence
of operations:

\begin{table}[ht]\caption{Numerical values of the six parameters
$a,$ $b,$ $\gamma,$ $\delta,$ $c,$ $d$ defining our formal
modelling (\ref{Eq:M}) of the quark mass function
$M(\underline{k}^2)$ in the chiral limit as retrieved from Fig.~2
of Ref.~\cite{PM97b}.}\label{Tab:M}
\begin{center}\begin{tabular}{lrrrrrr}\hline\hline\\[-1.5ex]
Parameter&\multicolumn{1}{c}{$a\left[\mbox{GeV}\right]$}
&\multicolumn{1}{c}{$b\left[\mbox{GeV}^2\right]$}
&\multicolumn{1}{c}{$\gamma$}&\multicolumn{1}{c}{$\delta$}
&\multicolumn{1}{c}{$c\left[\mbox{GeV}\right]$}
&\multicolumn{1}{c}{$d\left[\mbox{GeV}^{-2}\right]$}\\[1.5ex]\hline
\\[-1.5ex]Value&0.0705896&0.486542&1.48&0.75188&0.707983&1.52616
\\[1.5ex]\hline\hline\end{tabular}\end{center}\end{table}

\begin{enumerate}\item In order to get a firm grip on our inversion
problem under consideration, we first~need to convert the input
information, provided in pointwise shape by Fig.~2 of
Ref.~\cite{PM97b}, to a mathematical expression. We may achieve
this goal by construction of a convenient parametrization of the
quark mass function $M(\underline{k}^2)$ in terms of a very small
number of elementary (and hence easy-to-handle) functions. We
employ a six-parameter~ansatz:\pagebreak\begin{equation}
M(\underline{k}^2)=\frac{a}
{\left[1+\left(\underline{k}^2/b\right)^\gamma\right]^\delta}
+c\,\exp\!\left(-d\,\underline{k}^2\right).\label{Eq:M}
\end{equation}Fitting our parametrization (\ref{Eq:M}) of the
quark mass function, $M(\underline{k}^2),$ to the graph of its
momentum dependence in the chiral limit given in Ref.~\cite{PM97b}
provides, for the involved six parameters, $a,$ $b,$ $\gamma,$
$\delta,$ $c,$ and $d,$ the numerical values listed in
Table~\ref{Tab:M}. Accordingly, for lightlike momenta
($\underline{k}^2=0$), $M(\underline{k}^2)$ takes the value
$M(0)=a+c=0.778572\;\mbox{GeV}.$ Figure~\ref{Fig:M(p2)} depicts
our view (\ref{Eq:M}) on $M(\underline{k}^2)$ for both
double-logarithmic and linear scales.

\begin{figure}[hbt]\begin{center}\begin{tabular}{cc}
\psfig{figure=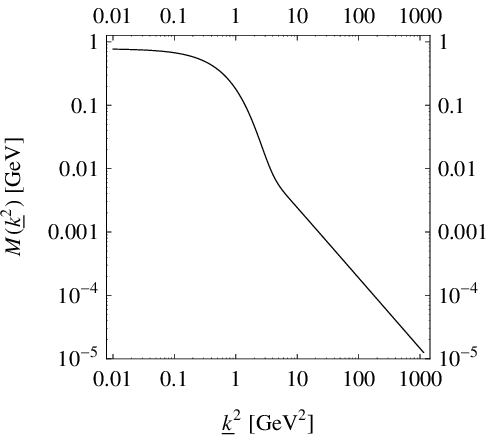,scale=1.599}&
\psfig{figure=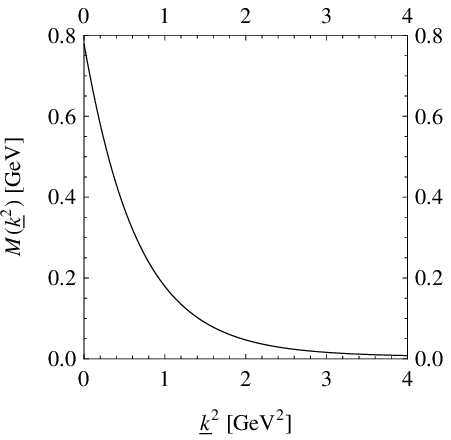,scale=1.599}\\(a)&(b)\end{tabular}
\caption{Mass function $M(\underline{k}^2),$ derived by solution
of the renormalization-group-improved rainbow-ladder approximation
to the Dyson--Schwinger equation for the quark propagator,
$S(\underline{k}),$ in the chiral limit (Fig.~2 of
Ref.~\cite{PM97b}), for double-logarithmic (a) and linear (b)
scales.} \label{Fig:M(p2)}\end{center}\end{figure}

\begin{table}[ht]\caption{Numerical values of the four parameters
($a,b,c,\gamma$) controlling the behaviour (\ref{Eq:Fp4}) of the
(sole) component $\varphi_2(p)$ of the Salpeter amplitude
(\ref{Eq:PSA}) of a spin-singlet quark--antiquark bound state
fixed by insertion of our parametrization (\ref{Eq:M}) of the
quark mass~function $M(\underline{k}^2)$ into the
``starting-point'' relationship (\ref{Eq:SP}) and subsequent
integration (\ref{Eq:SA}) with respect to~$\underline{k}_4.$}
\label{Tab:SCmom4}\begin{center}
\begin{tabular}{lrrrr}\hline\hline\\[-1.5ex]Parameter
&\multicolumn{1}{c}{$a\left[\mbox{GeV}^{2\,\gamma-\frac{3}{2}}\right]$}
&\multicolumn{1}{c}{$b\left[\mbox{GeV}\right]$}
&\multicolumn{1}{c}{$c\left[\mbox{GeV}^{-2}\right]$}
&\multicolumn{1}{c}{$\gamma$}\\[1.5ex]\hline\\[-1.5ex]
Value&8.7344&1.23635&2.57541&1.77044\\[1.5ex]\hline\hline
\end{tabular}\end{center}\end{table}

\item Inserting $M(\underline{k}^2)$ in form of its parametrization
(\ref{Eq:M}) into our starting-point relation (\ref{Eq:SP}) takes
us to the Goldstone-boson Bethe--Salpeter amplitude
$\Phi(\underline{k},0).$ Mimicking the $p_0$ integration requested
by the definition (\ref{Eq:SA}) by an integration of
$\Phi(\underline{k},0)$ with respect to $\underline{k}_4$ yields
the Salpeter amplitude $\phi(\bm{p}).$ For the outcome of this
\emph{numerical\/}~integration, the function $\varphi_2(\bm{p})$
multiplying the Dirac matrix $\gamma_5,$ a nearly perfect fit may
be found,\begin{equation}\varphi_2(p)=\frac{a}
{\left(b^2+p^2+c\,p^4\right)^\gamma}\ ,\qquad\|\varphi_2\|^2
\equiv\int_0^\infty{\rm d}p\,p^2\,|\varphi_2(p)|^2=1\
,\label{Eq:Fp4}\end{equation}with the numerical values of its just
four parameters $a,$ $b,$ $c,$ and $\gamma$ revealed in
Table~\ref{Tab:SCmom4}. Figure \ref{Fig:perfit} confronts, for the
Salpeter amplitude $\varphi_2(p),$ its parametrization
(\ref{Eq:Fp4}) with the direct output of the integration
(\ref{Eq:SA}). The maximum error is less
than~$0.0036\;\mbox{GeV}^{-3/2}.$

\begin{figure}[hbt]\begin{center}
\psfig{figure=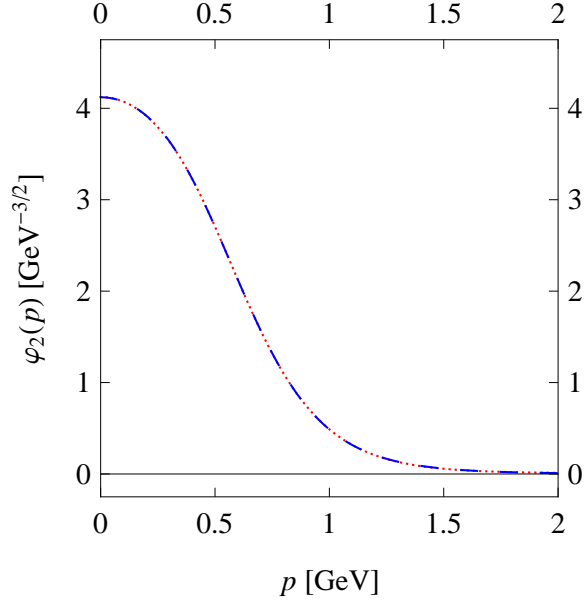,scale=1.7205}\caption{Comparison of
the result of a numerical integration, according to definition
(\ref{Eq:SA}), of the Salpeter amplitude $\Phi(\underline{k},0)$
emerging from the quark mass function (\ref{Eq:M}) (red~dotted
line) with our parametrization (\ref{Eq:Fp4}) defined by the
parameter values of Table~\ref{Tab:SCmom4} (blue dashed~line).}
\label{Fig:perfit}\end{center}\end{figure}

\item\label{Step:k2} Clearly, for the envisaged extraction of any
underlying interaction potential $V(r),$ we next have to move to
configuration space. It goes without saying that, in the~course of
performing the required Fourier transformations, we would like to
maintain sufficient control about the actual reliability of our
inversion technique. We may feel entitled to claim to have
achieved such goal if we are able to estimate the accuracy of our
results. To this end, we would like to perform the remaining
intermediate steps of the present inversion approach by as far as
reasonable analytical means. Consequently,~in~spite~of having at
our disposal, in form of Eq.~(\ref{Eq:Fp4}), a parametrization
pretty close to optimum, we prefer to proceed by use of an
approximation which is of slightly minor quality but resembles the
single-particle kinetic energy $E(p)$ sufficiently to provide, for
particular values of the quark mass, analytic access to $V(r).$
Hence, we continue with the ansatz
\begin{equation}\varphi_2(p)\approx\sqrt{\frac{\Gamma(2\,\gamma)}
{\sqrt{\pi}\,\Gamma(2\,\gamma-\frac{3}{2})}}\,
\frac{2\,b^{2\,\gamma-\frac{3}{2}}}{\left(p^2+b^2\right)^\gamma}\
,\qquad\|\varphi_2\|^2\equiv\int_0^\infty{\rm
d}p\,p^2\,|\varphi_2(p)|^2 =1\ ,\label{Eq:Fp}\end{equation}for
$\gamma>\frac{3}{4},$ with the appropriate values of the two
parameters $b$ and $\gamma$ given in~Table~\ref{Tab:SCmom}.

\begin{table}[ht]\caption{Numerical values of the (just two)
parameters $b,\gamma$ determining our ``user-friendly''
approximate parametrization (\ref{Eq:Fp}) of the single component
$\varphi_2(p)$ of the Salpeter amplitude (\ref{Eq:PSA}) of
spin-singlet quark--antiquark bound states derived from the quark
mass function~(\ref{Eq:M}).}\label{Tab:SCmom}
\begin{center}\begin{tabular}{lrr}\hline\hline\\[-1.5ex]
Parameter&\multicolumn{1}{c}{$b\left[\mbox{GeV}\right]$}
&\multicolumn{1}{c}{$\gamma$}\\[1.5ex]\hline
\\[-1.5ex]Value&1.69334&6.49292\\[1.5ex]\hline\hline
\end{tabular}\end{center}\end{table}

The proximity of convenient fit (\ref{Eq:Fp}) to ``perfect'' fit
(\ref{Eq:Fp4}) may be judged from Fig.~\ref{Fig:DSE_SC}(a).

\begin{figure}[hbt]\begin{center}\begin{tabular}{cc}
\psfig{figure=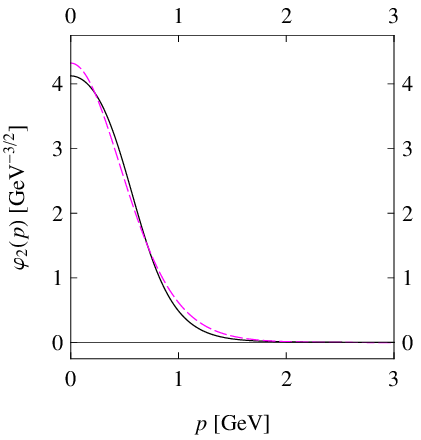,scale=1.7205}&
\psfig{figure=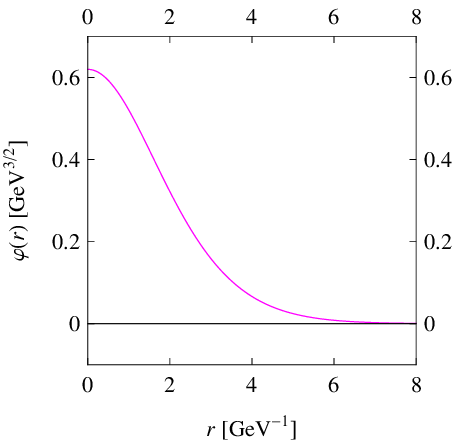,scale=1.7205}\\(a)&(b)\end{tabular}
\caption{Normalized radial independent Salpeter component
defining, at least for Salpeter equations (\ref{Eq:SE})
characterized by the Lorentz structure
$2\,\Gamma\otimes\Gamma=\gamma_\mu\otimes\gamma^\mu
+\gamma_5\otimes\gamma_5-1\otimes1$ of their interaction kernels,
the Salpeter amplitude (\ref{Eq:PSA}) of any spin-singlet state,
thus, in particular, of any pseudoscalar meson. In momentum-space
representation (a) we show its exact shape computed, via the
relationship (\ref{Eq:SP}), from the quark mass function of
Fig.~\ref{Fig:M(p2)} (black solid line) as well as an
approximation deduced from a fit to the simplified functional
dependence~(\ref{Eq:Fp}) (magenta dashed line) whereas in
configuration-space representation (b) we only depict the
behaviour of the Fourier--Bessel transform of the latter
approximation~(magenta solid~line).}\label{Fig:DSE_SC}\end{center}
\end{figure}

For the marginally simpler parametrization (\ref{Eq:Fp}), it is
then straightforward to obtain, by Fourier--Bessel transformation
of $\varphi_2(p),$ the Salpeter component $\varphi(r)$
analytically, in terms of the modified Bessel functions
$K_\eta(z)$ of the second kind of order $\eta\in{\mathbb R}$
\cite{AS}:$$\varphi(r)=\sqrt{\frac{\Gamma(2\,\gamma)}
{\sqrt{\pi}\,\Gamma(2\,\gamma-\frac{3}{2})}}\,
\frac{2^{2-\gamma}\,b^\gamma}{\Gamma(\gamma)}\,
r^{\gamma-\frac{3}{2}}\,K_{\frac{3}{2}-\gamma}(b\,r)\
,\qquad\|\varphi\|^2\equiv\int_0^\infty{\rm
d}r\,r^2\,|\varphi(r)|^2=1\ .$$Figure~\ref{Fig:DSE_SC}(b)
illustrates this configuration-space behaviour of our Salpeter
component.\item In general, the kinetic term's Fourier--Bessel
transform $T(r)$ can be computed merely by a numerical
integration. Thereafter, according to Eq.~(\ref{Eq:Po}), this
inversion procedure is easily completed by computing the sought
potential $V(r)$ by dividing $T(r)$ by $\varphi(r).$ The behaviour
of the resulting potentials for various quark masses is shown in
Fig.~\ref{Fig:V}.

\begin{figure}[hbt]\begin{center}\psfig{figure=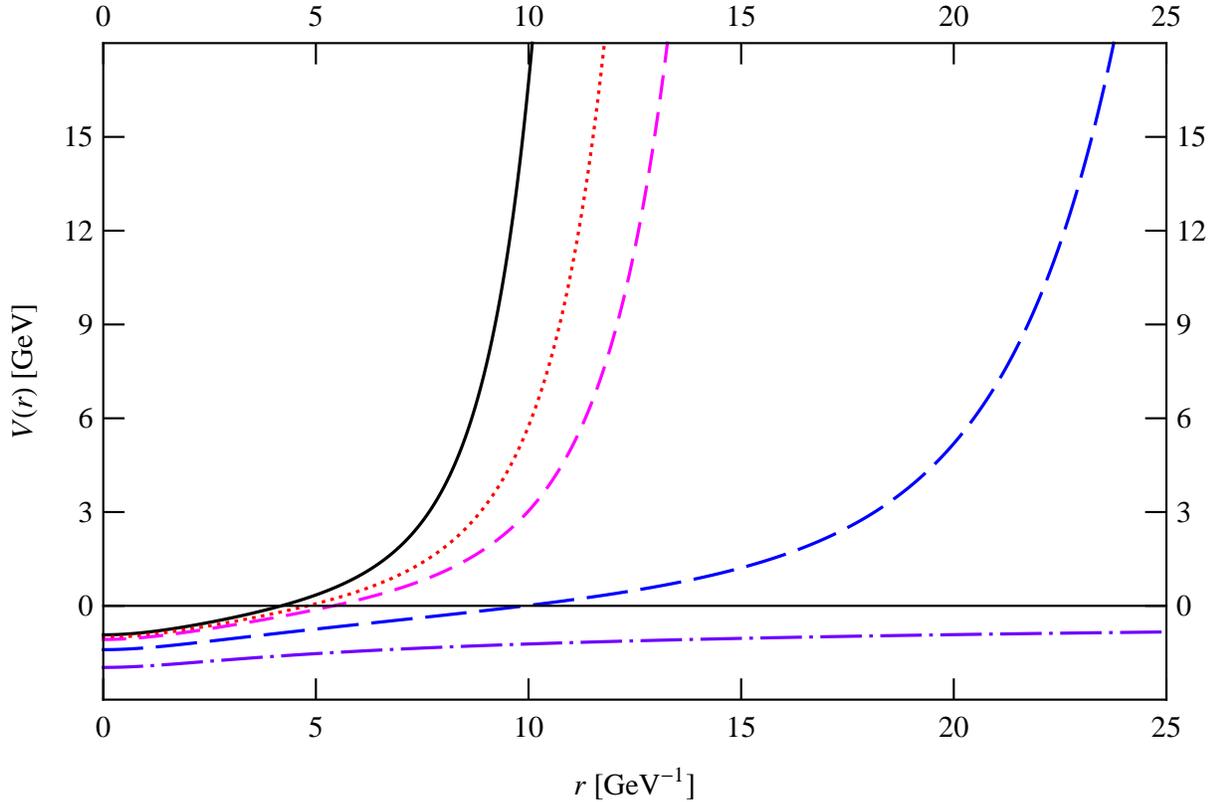,scale=1.7463}
\caption{Configuration-space potential $V(r)$ extracted from the
Salpeter equation (\ref{Eq:SE}) with interaction-kernel Lorentz
structure $2\,\Gamma\otimes\Gamma=\gamma_\mu\otimes\gamma^\mu
+\gamma_5\otimes\gamma_5-1\otimes1,$ for typical quark-mass value,
$m$: $m=0$ (black solid line), $m=0.35\;\mbox{GeV}$ (red dotted
line), $m=0.5\;\mbox{GeV}$~(magenta short-dashed line),
$m=1\;\mbox{GeV}$ (blue long-dashed line), and $m=b$ (violet
dot-dashed line).}\label{Fig:V}\end{center}\end{figure}

As promised when motivating our step \ref{Step:k2}, owing to the
pivotal simplification (\ref{Eq:Fp}), for, at least, two distinct
choices of the quark mass $m$ the Fourier--Bessel transform of the
kinetic term $E(p)\,\varphi_2(p),$ and thus the potential $V(r),$
can be found in analytical~form:\begin{enumerate}\item If that
common quark mass $m$ vanishes, that is, for $m=0,$ the kinetic
term $T(r)$ may be formulated by use of the modified Bessel
functions of the first~kind $I_\eta(z),$ of order $\eta\in{\mathbb
R}$ \cite{AS}, and the modified Struve functions ${\bf
L}_\eta(z),$ of order~$\eta\in{\mathbb R}$ \cite{AS}:\pagebreak
\begin{align*}T(r)&=\sqrt{\frac{\pi^\frac{3}{2}\,\Gamma(2\,\gamma)}
{\Gamma(2\,\gamma-\frac{3}{2})}}\,\frac{2^{1-\gamma}}
{\sin(\gamma\,\pi)\,\Gamma(\gamma)}\,\frac{1}{r^{\frac{5}{2}}}\\&\times
\left[2\,(1-\gamma)\,(b\,r)^\gamma\,I_{\gamma-\frac{3}{2}}(b\,r)
-(b\,r)^{1+\gamma}\,I_{\gamma-\frac{1}{2}}(b\,r)\right.\\
&\!\left.{}+(b\,r)^\gamma\,{\bf L}_{\frac{3}{2}-\gamma}(b\,r)
+(b\,r)^{1+\gamma}\,{\bf L}_{\frac{5}{2}-\gamma}(b\,r)
+\frac{\sin(\gamma\,\pi)}{2^{\frac{3}{2}-\gamma}\,\pi^\frac{3}{2}}\,
\Gamma(\gamma-2)\,(b\,r)^{\frac{5}{2}}\right].\end{align*}We do
not see a genuine need for reproducing here the rather lengthy
expression of $V(r)$ that arises upon division of the above $T(r)$
by $\varphi(r)$: see the solid curve in Fig.~\ref{Fig:V}. With
this explicit expression for $V(r)$ at hand, it is trivial to
characterize the behaviour of $V(r)$ in the limit of either small
or large interquark distances~$r$:
\begin{align*}V(0)&=-\frac{2\,b\,\sqrt{\pi}}
{\sin(\gamma\,\pi)\,\Gamma(3-\gamma)\,\Gamma(\gamma-\frac{3}{2})}
\qquad\mbox{for}\ \gamma>2\ ,\\V(r)&\xrightarrow[r\to\infty]{}
\frac{\gamma\,(\gamma-1)}{4\,\sin(\gamma\,\pi)}\,
\frac{\exp(2\,b\,r)}{r}\ .\end{align*}Thus, for appropriate values
of $\gamma,$ we get a \emph{confinement-betraying\/} rise to
infinity:$$V(r)\xrightarrow[r\to\infty]{}+\infty\qquad\mbox{for}\
\gamma\in(2\,n,2\,n+1)\ ,\quad n=1,2,3,\dots\ .$$By formula
(6.1.17) of Ref.~\cite{AS}, the expression
$\sin(\gamma\,\pi)\,\Gamma(3-\gamma)$ is \emph{nonvanishing},
$$\sin(\gamma\,\pi)\,\Gamma(3-\gamma)=\frac{\pi}{\Gamma(\gamma-2)}
>0\qquad\mbox{for}\ \gamma>2\ ;$$accordingly, at spatial origin
$r=0,$ the potential $V(r)$ assumes a finite value:~for the
parameters of Table~\ref{Tab:SCmom}, $V(0)=-0.926813\;\mbox{GeV}$
(cf.\ the solid curve in Fig.~\ref{Fig:V}).\item If, by chance,
the common quark mass $m$ is exactly equal to the mass parameter
$b$ in our ansatz (\ref{Eq:Fp}), that is, for $m=b,$ the kinetic
term $T(r)$ turns out to~involve only the modified Bessel
functions of the second kind $K_\eta(z),$ of order
$\eta\in{\mathbb R}$ \cite{AS}:
$$T(r)=\sqrt{\frac{\Gamma(2\,\gamma)}
{\sqrt{\pi}\,\Gamma(2\,\gamma-\frac{3}{2})}}\,
\frac{2^{\frac{5}{2}-\gamma}\,b^{\frac{1}{2}+\gamma}}
{\Gamma(\gamma-\frac{1}{2})}\,r^{\gamma-2}\,K_{2-\gamma}(b\,r)\
.$$With the overall normalization dropping out, the emerging
potential~$V(r)$~reads
$$V(r)=-\frac{\Gamma(\gamma)}{\Gamma(\gamma-\frac{1}{2})}\,
\sqrt{\frac{2\,b}{r}}\,\frac{K_{2-\gamma}(b\,r)}
{K_{\frac{3}{2}-\gamma}(b\,r)}\ .$$This behaviour of $V(r)$ is
reflected graphically by the dot-dashed curve in~Fig.~\ref{Fig:V}.
For such simple $V(r)$ shape, its $r\to 0$ and $r\to\infty$ limits
can be directly read~off:
\begin{align*}V(0)&=-\frac{b\,\Gamma(\gamma-2)\,\Gamma(\gamma)}
{\Gamma(\gamma-\frac{3}{2})\,\Gamma(\gamma-\frac{1}{2})}
\qquad\mbox{for}\ \gamma>2\ ,\\V(r)&\xrightarrow[r\to\infty]{}
-\frac{\Gamma(\gamma)}{\Gamma(\gamma-\frac{1}{2})}\,
\sqrt{\frac{2\,b}{r}}\xrightarrow[r\to\infty]{}0\ .\end{align*}
Thus, for rising separation $r, $ this potential starts from a
finite, negative value~at the origin $r=0,$ namely,
$V(0)=-1.96928\;\mbox{GeV}$ for the two parameter values~of
Table~\ref{Tab:SCmom}, but stays below zero even for $r\to\infty;$
cf.\ the dot-dashed curve in Fig.~\ref{Fig:V}.\end{enumerate}In
general, the Fourier--Bessel transform of the kinetic term,
$T(r),$ and, consequently, the resulting potential, $V(r),$ have
to be computed by numerical integration. \pagebreak
Figure~\ref{Fig:V} presents the outcomes of such undertaking, for
a couple of selected quark masses $m$ in the interval $0\le m\le
b.$ Needless to say that, in both analytically accessible
cases,~our numerical results exhibit perfect agreement with the
above explicit findings for $V(r).$\end{enumerate}Numerically, the
\emph{oscillatory\/} behaviour induced by the very definition of
the Fourier--Bessel transformation requires in its application a
particularly careful analysis. In view~of this, the possibility of
the independent verification of one's numerical findings, for
exceptional cases, by corresponding analytic results should be
considered as a precious and welcome~bonanza.

\section{Summary, Conclusion, Interpretation and Outlook}
\label{Sec:SCIO}Abandoning for the moment most of our previous
ambitions towards analytic treatments~of the lightest pseudoscalar
mesons within the formalism of the instantaneous Bethe--Salpeter
approach \cite{WL13,WL15,WL16:ARP}, in the present analysis we
derived, for the Salpeter equation (\ref{Eq:SE}), the form of that
interaction kernel which is capable of describing Goldstone-type
pseudoscalar~mesons, from the quark mass function, rendered
accessible by a fundamental relationship (resulting from the
chiral symmetry of QCD \cite{PM97a,PM97b}), between the quark
propagator, on the one hand, and the meson Bethe--Salpeter
amplitude, on the other hand. For increasing effective quark mass
$m,$ the gross behaviour of $V(r)$ resembles that observed in
Ref.~\cite{WL15}: a rise to infinity for sufficiently small $m$
but an approach to a finite nonpositive value for larger $m.$ This
is in full accordance with the inevitable large-$m$ behaviour of
$V(r)$ already demonstrated~in~Ref.~\cite{WL15}:
$$E(p)\xrightarrow[m\to\infty]{}m\qquad\Longrightarrow\qquad
T(r)\xrightarrow[m\to\infty]{}m\,\varphi(r)\qquad\Longrightarrow
\qquad V(r)\xrightarrow[m\to\infty]{}-m\ .$$

The origin of any such qualitatively different behaviour of the
derived potentials $V(r)$ is easily identified: For given mass $m$
of the bound-state constituents, the configuration-space
quantities $T(r)$ and $\varphi(r),$ as mere Fourier--Bessel
transforms, and thus likewise their ratios, the potentials
$V(r)=-T(r)/\varphi(r),$ are unambiguously determined already~by
the Salpeter component $\varphi_2(p)$ forming the exclusive
momentum-space input of our inversion procedure. Thus, any
substantial differences of the predictions for $V(r)$ have to be
attributed to~$\varphi_2(p).$ For the sake of comparison, let us
represent the approximate behaviour of $\varphi_2(p)$ in the~form
$$\varphi_2(p)\propto\frac{1}{\left(p^2+\mu^2\right)^\nu}\ ,$$with
some characteristic mass scale, $\mu,$ and discuss the associated
exponents $\nu$ encountered in this and two previous analyses: In
Ref.~\cite{WL15}, the momentum dependence of the chiral-limit
quark mass function in the ultraviolet limit of large spacelike
momenta, deduced on general grounds \cite{PM97b}, implied
$\nu=\frac{3}{2}.$ In Ref.~\cite{WL16:ARP}, we modelled the
(confinement-promoting) presence of an inflection point in the
quark mass function by an admixture that entailed a somewhat
modified value of the effective $\nu,$ depending on the relative
amount of this admixture. Here, this exponent is, of course, equal
to the numerical value of the parameter $\gamma$ in the simplified
parametrization (\ref{Eq:Fp}), $\nu=\gamma\approx\frac{13}{2}$
according to Table \ref{Tab:SCmom}, and markedly~larger,~because
in the simplified form (\ref{Eq:Fp}) this parameter must take care
of the $p^4$ term in the parametrization (\ref{Eq:Fp4}) which,
otherwise, would dominate the behaviour of $\varphi_2(p)$ at large
$p.$ This is reflected by the nonsingular behaviour of
$\varphi(r)$ at the origin, whence there is no need for a
counterbalancing singularity of $V(r)$ at the origin that allows
for the sought masslessness of the bound states.

So far, we have harvested our (analytical or numerical) input in
Euclidean space, i.e., at spacelike quark momenta. Beyond doubt,
it will be interesting to move to Minkowski~space and to exploit
findings for timelike quark momenta, to the extent such results
are available.

\end{document}